\documentclass[letterpaper]{iopart}
\usepackage{iopams}
\jl{6}

\input epsf

\def\be {\begin{equation}}
\def\ee {\end{equation}  }
\def\beq{\begin{eqnarray}}
\def\eeq{\end{eqnarray}  }

\def\lp{\left(}
\def\rp{\right)}
\def\lb{\left[}
\def\rb{\right]}

\def\sq{^2}
\def\PD{\partial}

\def\rarrow{\rightarrow}

\def\sp{_{\rm sp}}
\def\dn{_{\rm dn}}

\def\poundcode{\char`\$}
\def\pound{\it\poundcode}
\def\LSODE{{\tt LSODE}}
\def\M{{\cal M}}
\def\B{{\cal B}}

\begin{document}

\title{Critical phenomena in perfect fluids}

\author{David W. Neilsen and Matthew W. Choptuik}
\address{Center for Relativity,
         The University of Texas at Austin,
         Austin, TX 78712-1081}

\begin{abstract}
We investigate the gravitational collapse of a
spherically symmetric, perfect fluid
with equation of state $P = (\Gamma -1)\rho$. We restrict
attention to the ultrarelativistic (``kinetic-energy-dominated'', 
``scale-free'') limit where black hole formation is anticipated to turn on at 
infinitesimal black hole mass (Type~II behavior).
Critical solutions (those which sit at the threshold of 
black hole formation in parametrized families of 
collapse) are found by solving the system of
ODEs which result from a self-similar {\em ansatz,}
{\it and} by solving the full Einstein/fluid PDEs in spherical symmetry.
These latter PDE solutions (``simulations'') extend the pioneering 
work of Evans and Coleman
($\Gamma = 4/3$) and verify that the continuously
self-similar solutions previously found by Maison and Hara \etal
for $1.05 \le \Gamma  \lesssim 1.89$ {\em are} (locally) unique 
critical solutions.
In addition, we find strong evidence that globally regular critical
solutions {\it do} exist for $1.89 \lesssim \Gamma \le 2$, 
that the sonic point for $\Gamma\dn \simeq 1.8896244$
is a degenerate node, and that the sonic points for $\Gamma >\Gamma\dn$
are nodal points, rather than focal points  as previously reported.
We also find a critical solution for $\Gamma=2$, and 
present evidence that it is continuously self-similar
and Type~II. Mass-scaling exponents for all of the critical solutions 
are calculated by evolving near-critical initial data, with results 
which confirm and extend previous calculations based on linear 
perturbation theory.  Finally, we comment on critical solutions 
generated with an ideal-gas equation of state.  
\end{abstract}


\submitted

\vskip 2pc

\section{Introduction}
\label{sec:introduction}

The formation of black holes is an exciting topic in general
relativity, and a class of solutions which exists precisely
at the threshold of black hole formation has recently generated considerable
interest.  These solutions have surprising properties,
reminiscent of some thermodynamic systems near phase 
transitions, and, by analogy, have been called critical solutions. 
Critical phenomena
in gravitational collapse were first discovered empirically in simulations
of the massless Klein-Gordon field minimally coupled to gravity 
(EMKG)~\cite{mwc93}.
Subsequent studies have shown that critical behavior is present
in a variety of physical systems, and indicate that the phenomena
are generic features of gravitational collapse in general relativity.
In this paper we focus on the critical solutions for a spherically
symmetric perfect fluid with equation of state $P=(\Gamma-1)\rho$,
where $\rho$ is the total energy density and $\Gamma$ is constant,
and present new solutions for $\Gamma \gtrsim 1.89$.   While we 
provide a brief introduction to critical solutions, the review
by Gundlach is an excellent introduction to critical phenomena~\cite{cg97}, 
and additional information can be found in~\cite{mwc98}.

\subsection{Basic properties of critical solutions}

Imagine an experiment to investigate the details of gravitational collapse and
black hole formation by imploding shells of fluid with a fixed equation
of state. 
The initial energy density in the shell might be
\be
\rho = A_0\exp\lb -\lp r - r_0 \rp^2 / \Delta^2\rb,
\ee
where $A_0$, $r_0$, and $\Delta$ are parameters.
In the course of the experiment we fix two of the three parameters,
and allow only one of them, which we label $p$, to vary.  
For small $p$ (assuming the fluid's initial kinetic energy
is sufficiently large---i.e. in what one might call the 
ultrarelativistic limit), the fluid implodes through the origin and
completely disperses.  However, for $p$ sufficiently large,
in particular for $p$ larger than some critical value $p^\star$,
a black hole forms during the implosion, trapping some of the 
matter/energy within a finite radius.
In the exactly critical case,
$p=p^\star$, which represents the threshold of black hole formation,
the evolution temporarily asymptotes to a special critical 
solution, $Z^\star$,
which has a number of interesting properties, including scale invariance
(self-similarity) and universality.  
The critical solution is universal in the sense
that if we now use  different ``interpolating families'' to probe
the threshold of black hole formation, we will generically 
find the same critical solution (provided we remain in the 
ultrarelativistic regime).  Additionally, in the super-critical
regime $p>p^\star$, the black hole masses are well described by
a scaling law
\be
M_{{\rm BH}}(p) \propto \left| p - p^\star \right|^\gamma.
\label{eq:bhmass}
\ee
Here the mass-scaling exponent $\gamma$ is {\em also} universal in the
sense that it is independent of the particular choice of initial data
family.  (However, as first predicted by Maison~\cite{dm}, and 
Hara, Koike and Adachi~\cite{hka, kha2},
and as discussed in detail below,
$\gamma$ {\em is} a function of the adiabatic index $\Gamma$.)  

One of the most profound consequences of the self-similar nature 
of critical collapse is that black hole formation in the 
ultrarelativistic limit turns on at {\em infinitesimal} 
mass.  In analogy with second-order phase transitions in 
statistical mechanics, we refer to this behavior as 
Type~II.  As we will discuss shortly, Type~I behavior,
wherein black hole formation turns on with {\em finite} mass 
in interpolating families, has also been seen in various models of 
collapse and it is undoubtedly present in at least some of the perfect 
fluid models considered here.

\subsection{Critical solutions and one-mode instability}

A crucial feature of the critical solutions sketched above is 
that they are, {\em by construction}, unstable.  If this is not 
obvious, one should observe that the critical solution is 
{\em not} a long-time ($t\to\infty$) solution of the equations 
of motion.  Indeed, as sketched above, the only long-time stable 
``states'' one finds from evolutions of a generic 
ultrarelativistic family of initial data either have 
(i) all of the fluid dissipated to arbitrarily large radii, with
 (essentially) flat spacetime in the interior, or (ii)
some fluid dissipated to arbitrarily large radii, with a black-hole
in the interior.  The critical solution, $Z^\star$, on the other hand, 
exists just at the threshold of black hole formation, and, in
near-critical evolutions, the dynamics asymptotes to $Z^\star$ {\em only}
during the strong-field dynamical epoch.   For any given 
initial data, this strong-field regime persists for a 
{\em finite} amount of time (as measured, for 
example, by an observer at infinity).  Eventually (and in fact, on 
a dynamical time scale) any non-critical data will evolve into 
one of the two stable end states. 

Although the unstable nature of critical solutions was clear 
from the earliest phenomenological studies, considerable 
insight has been gained from the observation by Koike, Hara 
and Adachi~\cite{kha95}
that the ``sharpness'' of the critical behavior seen in Type~II 
collapse suggests that the critical solutions have {\em exactly one} unstable 
mode in perturbation theory.   This {\em ansatz} immediately explains 
the universality of the critical solution:  as $p \to p^\star$, one 
is effectively directly tuning out the single unstable mode from 
the initial data.  Furthermore, using the self-similarity of the 
dynamics in the near-critical regime and a little dimensional 
analysis, it is an easy matter to relate the mass-scaling exponent 
to the Lyapunov exponent associated with the single mode.  In fact,
since the pioneering work by Koike \etal, this picture of 
Type~II critical solutions as one-mode unstable, self-similar
``intermediate attractors'' has been validated for essentially every 
spherically-symmetric model where Type~II behavior has been observed
in the solution of the full equations of motion.

Moreover, the perturbative analysis applies equally well to 
Type~I critical solutions which, arguably, have been well known 
to relativists and astrophysicists for decades, although perhaps 
not in the context of interpolating families.  In this case, the 
critical solution is an unstable {\em static} or {\em periodic} 
configuration which, depending on how it is perturbed, will
either completely disperse, or collapse to a finite-mass black hole. 
Once again, one generically finds that such solutions have a 
{\em single} unstable eigenmode, whose Lyapunov exponent is now
a measure of the increase in lifetime of the unstable configuration
as one tunes $p \to p^\star$.   Type~I behavior has been observed 
in the collapse of Yang-Mills \cite{mwc_ym} 
and massive scalar fields \cite{pb_msf}, and, as 
mentioned above, there is every reason to expect that it will occur 
in perfect fluid collapse.   Indeed, it is well known that 
relativistic stellar models often exhibit a turn-over in total 
mass as a function of central density.  Stars past this turn-over 
(generalized Chandrasekhar mass) are well known to be unstable, 
and, in fact, are almost certainly Type~I solutions.

\begin{figure}
\epsfxsize = 8cm
\centerline{\epsffile{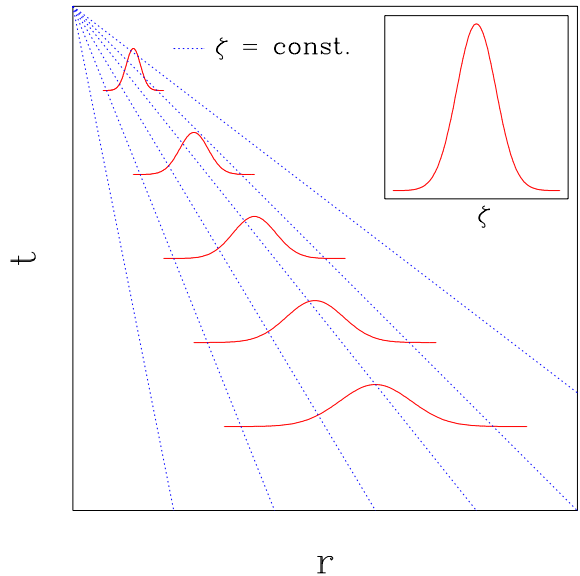}}
\caption{A schematic diagram showing a continuously self-similar (CSS) pulse
at five different times as it moves toward the origin $r=0$.
The dotted lines
are lines of constant $\zeta = r/t$, the similarity variable.
These lines converge at the space-time origin $(r,t) = (0,0)$ in
the upper left-hand corner of the plot, and the
inset shows the pulse as a function of $\zeta$.
As the pulse moves toward the
origin, it appears the same on smaller and smaller length scales.
}
\label{fig:cs}
\end{figure}

\subsection{Critical solutions and self-similarity}

Heuristically, systems exhibiting self-similarity appear identical
over many different spatial and/or temporal scales, and generally arise 
in physical situations in which there are no natural length scales.  
Here it is important to note that a scale-free solution can be generated 
from a model which {\em does} have specific length scales, provided 
that the scaling solution represents a ``self-consistent'' limit. 
The current case of fluid collapse provides a perfect example.  The 
rest mass of the fluid {\em  does} set a length-scale, but the 
Type~II critical solutions describe an ultrarelativistic limit 
wherein the rest-mass of the solution is irrelevant.  To put this 
another way, we can have {\em solutions} of the equations of motion 
which have greater symmetry (scale symmetry in this case) than
the equations of motion themselves.

Self-similarity can be either continuous (CSS) or discrete (DSS), and
both types have been observed in critical gravitational collapse.
The perfect fluid critical solutions have
continuous self-similarity of the {\it first} kind, a particularly
simple self-similarity wherein the solutions can be written
solely as functions of dimensionless variables, such as
$\zeta = r/t$, 
where $r$ is the radial coordinate in spherical symmetry and $t$ is the
coordinate time.  An example of a CSS function of the
first kind is shown schematically on a spacetime diagram in  \fref{fig:cs}. 

At this point we should also note that the self-similar 
nature of Type~II critical solutions provides 
a link between work on black-hole critical phenomena and the 
large body of literature dealing with self-similarity 
in the context of Einstein gravity (see Carr and Coley~\cite{cc1} 
for a recent, extensive review).   The self-similar {\em ansatz\/} has 
been widely employed, not only to 
produce more tractable problems, but also in investigations of 
possible mechanisms to generate counter-examples to the
cosmic censorship conjecture.  However, it is clear that 
not all self-similar solutions will be relevant to critical 
collapse, particularly if we restrict the definition of critical
to ``one-mode unstable''.   Moreover, because most studies 
which are based on the self-similar {\em ansatz} have only considered
the solutions themselves, and not perturbations thereof, it has 
proven non-trivial to identify which self-similar perfect-fluid 
solutions previously discussed in the literature are relevant 
to critical collapse~\cite{cc1,goliath,cc2}.

\subsection{Review of previous perfect-fluid studies}

Shortly after the discovery of Type~II behavior in scalar 
field collapse~\cite{mwc93}, Evans, having been frustrated in attempts to 
``analytically'' understand the massless-scalar critical
solutions, turned his attention 
to perfect-fluid collapse.  Armed with the intuition that 
the self-similarity of critical collapse was a defining 
characteristic, and aware of the existence of 
continuously self-similar relativistic fluid flows,
he and Coleman~\cite{cejc} considered collapse in the 
context of the specific equation of state,  
$P = \frac{1}{3}\rho$.  Significantly, they were able 
to construct a single critical solution, both from the self-similar 
{\em ansatz} (i.e. by solving ODEs), and by tuning the initial
data used in solution of the full partial differential
equations of motion.  Moreover, they noted that a perturbation analysis 
about the inherently unstable critical solution would provide an accurate
description of the near-critical dynamics, including the calculation
of the mass-scaling exponent $\gamma$.

Such a perturbation analysis was quickly carried out (again for the 
case $P = \frac{1}{3}\rho$) by Koike \etal~\cite{kha95}, who, as 
mentioned above, made the crucial additional observation that the 
``sharp'' transition in the mass scaling
suggested that there was only {\em one} growing unstable mode associated
with the critical solution, and that the 
Lyapunov exponent of the mode was simply the reciprocal of the 
mass-scaling exponent $\gamma$.  Their analysis fully 
validated this conjecture---in particular, they found strong evidence 
for a single unstable mode with Lyapunov exponent $2.81055\ldots\,$,
corresponding 
to $\gamma = 0.3558019\ldots\,$, 
in excellent agreement with Coleman and Evans' 
``measured'' value, $\gamma \approx 0.36$. 

At about the same time, Maison~\cite{dm}---assuming that the critical 
solutions would be continuously self-similar for other values of 
$\Gamma$---adopted the CSS {\em ansatz} for the more general
equation of state $P = (\Gamma - 1)\rho$.  He was able to 
construct CSS solutions for $1.01 \le \Gamma \le 1.888$, and 
additionally presented strong evidence that all of the solutions
were one-mode unstable.  Furthermore, the Lyapunov exponents,
and hence the mass-scaling exponents were found to be $\Gamma$-dependent,
with, $\gamma$, for example, varying from $\gamma = 0.1143$ for 
$\Gamma = 1.01$ to $\gamma = 0.8157$ for $\Gamma= 1.888$.  
These calculations were particularly notable for providing early evidence
that $\gamma$ was not a ``truly'' universal exponent, in the sense 
of having the same value across {\em all} possible models of 
collapse. 

One interesting outcome of Maison's linearized analysis
of the equations of motion about the sonic 
point, was that at $\Gamma \simeq 1.888$,
two of the eigenvalues of the linearized problem degenerated, and
the sonic point apparently changed  from a node to a focus.  This led 
Maison to conclude that regular self-similar solutions did not exist 
for $\Gamma \gtrsim 1.89$.
A similar analysis by Hara, Koike and Adachi~\cite{hka,kha2} (expanding 
on their previous work) again suggested a change in solution
behavior at $\Gamma \approx 1.89$.  Those authors computed
the CSS solutions, unstable eigenmodes, and eigenvalues for 
$1 < \Gamma \le 1.889$, with results essentially identical to 
Maison's.  
Evans and Perkins~\cite{perkins} 
also performed the linear stability analysis for $\Gamma\le 1.888$,
finding the same results reported by Maison.  In addition, they
performed the first critical solution searches using the full
set of PDEs for $1.05\le \Gamma \le 1.5$, confirming that the
CSS solutions are the unique critical solutions for $\Gamma$ is this range.
Goliath \etal \cite{goliath} discussed, in the 
wider context of timelike self-similar fluid solutions, the mode structure
of the $P = (\Gamma - 1)\rho$ CSS solutions, and reported that 
physical solutions do not exist for $\Gamma \gtrsim 1.89$.
More recently, after this paper appeared in pre-print form, Carr \etal
have extended this work for $\Gamma \gtrsim 1.89$~\cite{carr1,carr2}.

Furthermore, the conclusion of these linear 
perturbation analyses---that regular
critical solutions for $\Gamma \gtrsim 1.89$ do not exist---has inspired
various proposals regarding the nature of $\Gamma \gtrsim 1.89$ critical 
solutions~\cite{cg97,cc1,brady_cai}.
(Recall that as long as we can set up interpolating data,
there {\em will} be a critical solution, virtually by definition).
One proposal is that a loss of analyticity at the
sonic point for $\Gamma \gtrsim 1.89$
violates a condition required to find the ODE solutions.
Other proposals have suggested that the solution might become Type~I,
discretely self-similar, or display a mixture of DSS and CSS behavior.
Some of these conjectures were evidently based
on the fact that, under certain conditions,
a stiff ($P=\rho$) perfect fluid
can be formally identified with the EMKG system~\cite{taub,mad1,mad2}.  
For example, it has been conjectured~\cite{cg97,cc1,brady_cai} that 
at some point as $\Gamma \rightarrow 2$, the
critical solution might begin to display the discrete self-similarity 
characteristic of the EMKG critical solution.
Brady and Cai~\cite{brady_cai} have previously computed threshold 
solutions for $\Gamma \le 1.98$ using the full fluid equations of motion,
finding---in all cases examined---evidence that the critical solutions 
are both CSS and Type~II.
Using a two-step Lax-Wendroff numerical scheme to integrate the 
fluid equations, they calculated mass-scaling exponents by evolving
supercritical initial data.
However their code has severe resolution limitations,
being able to observe scaling only over two orders of magnitude in
$|p-p^\star |$.

Yet, lacking solutions for $\Gamma=2$, and high resolution solutions
near $\Gamma\simeq 1.89$, it was still expected that 
the perfect fluid critical solution changed its character
as $\Gamma\to 2$.
As we will discuss below, this does not seem to be the case, and in fact,
$\Gamma \simeq 1.89$ seems problematic only in the context of the the 
precisely self-similar {\em ansatz.}  Specifically, we have strong 
evidence that the CSS {\em ansatz} generates an increasingly 
ill-conditioned problem as $\Gamma \to 1.8896244 \ldots\,$, but that 
the PDEs remain perfectly well-behaved there.

\subsection{Summary of results}

Using a new perfect-fluid collapse 
code described in detail in~\cite{nc1}, this paper 
examines the perfect fluid critical solutions for the full range
of $\Gamma$, $1.05 \le \Gamma \le 2$, concentrating
on those solutions for $\Gamma \gtrsim 1.89$.
In addition to confirming the expected picture for 
$1.05 \le \Gamma \le 1.889$, we present further evidence that one-mode 
unstable CSS solutions exist in the regime $\Gamma \gtrsim 1.89$, up to
and including, the limiting case $\Gamma = 2$.  As part of this 
study, we are lead to re-investigate the CSS {\em ansatz,} and,
in particular, the apparent change in solution behavior for 
$\Gamma \simeq 1.889$.  In contrast to previous work, we solve 
the ODEs using arbitrary precision floating point arithmetic,
which proves crucial to the analysis leading to our 
claim that the sonic points for 
$\Gamma\ge \Gamma\dn \simeq 1.8896244$
are {\em nodal} points, rather than {\em focal} points.
We show that the CSS solutions for $\Gamma \gtrsim 1.89$
are qualitatively identical to those for $\Gamma \lesssim 1.89$, 
and verify that they can be computed from the full equations 
of motion (again using the code described in~\cite{nc1}). We 
also use simulations to compute mass scaling exponents in the 
regime $\Gamma \gtrsim 1.89$, where, interestingly, we find 
evidence to suggest that $\gamma \to 1$ as $\Gamma \to 2$.

Finally, researchers in this field have tacitly assumed that it 
suffices to consider $P = (\Gamma  - 1) \rho$ in the context 
of critical collapse, since (1) the critical solution naturally 
``drives itself'' to the ultrarelativistic (kinetic-energy-dominated)
regime  and (2)  $P = (\Gamma  - 1) \rho$  is the only
equation of state compatible with self-similarity~\cite{aotp,ce93}.  
We explicitly verify this assumption 
(albeit on a case-by-case basis) by performing 
computations---using an ideal-gas equation of state---which display 
the anticipated behavior.


\section{Construction of the precisely critical solutions}
\label{sec:odes}

Spherically symmetric
perfect fluid critical solutions can be constructed
by searching for globally regular, spherically symmetric,
CSS solutions to Einstein's 
equations \cite{cejc}.
As discussed in the Introduction, many 
studies of CSS perfect fluid spacetimes have appeared in 
the literature, and many properties of these
solutions, such as their behavior near sonic points, are well 
known~\cite{dm, hka, cc1, goliath, aotp, ob, bh1, bh2, tfrh}.
The self-similarity {\it ansatz} reduces the equations to first
order, autonomous ODEs, and this section focuses on the solution of these
ODEs for $\Gamma \gtrsim 1.89$.  We use the equations derived by 
Hara \etal~\cite{hka}, and
refer the reader to that paper for additional information
concerning the derivation of the equations, and basic methods for 
their solution.

\begin{figure}
\epsfxsize 9cm
\centerline{\epsffile{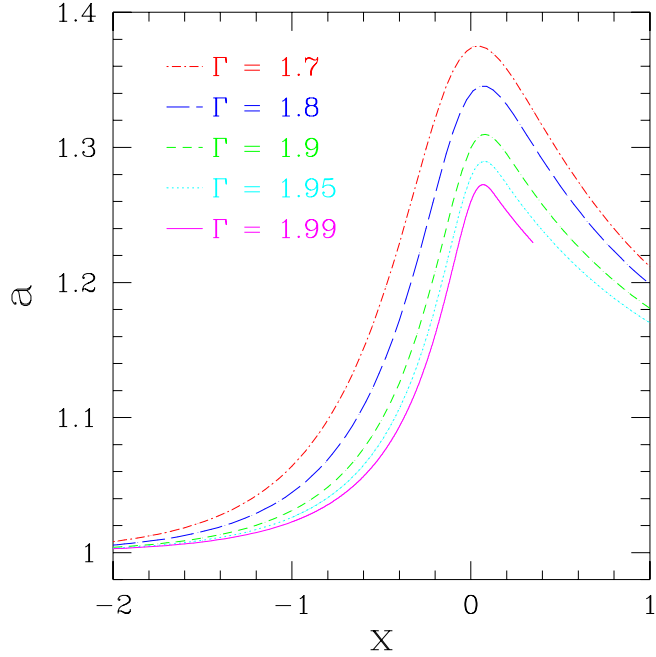}}
\caption{ The geometric variable $a$ of the critical 
solution for several values of $\Gamma$. The sonic point is at $x=0$.
The ODEs were integrated using the {\sl Maple V} implementation of \LSODE\ 
with 30 digits and an absolute error tolerance $\varepsilon = 10^{-18}$.
We are increasingly unable to integrate these solutions outwards as
$\Gamma\rightarrow 2$.  This often occurs owing to a loss of
numerical precision as $\omega \to 0$, and the Lorentz factor, 
$W = 1/\sqrt{1-v^2}$, 
becomes large (see \fref{fig:ec_w}).
}
\label{fig:ec_a}
\end{figure}

\begin{figure}
\epsfxsize 9cm
\centerline{\epsffile{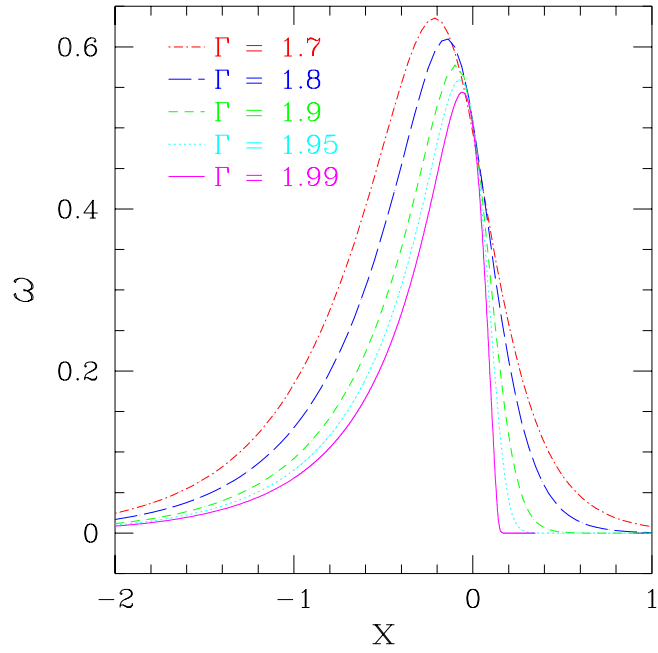}}
\caption{ The fluid variable $\omega$ of the critical
solution for several values of $\Gamma$.
The sonic point is at $x=0$.
}
\label{fig:ec_o}
\end{figure}

\begin{figure}
\epsfxsize 9cm
\centerline{\epsffile{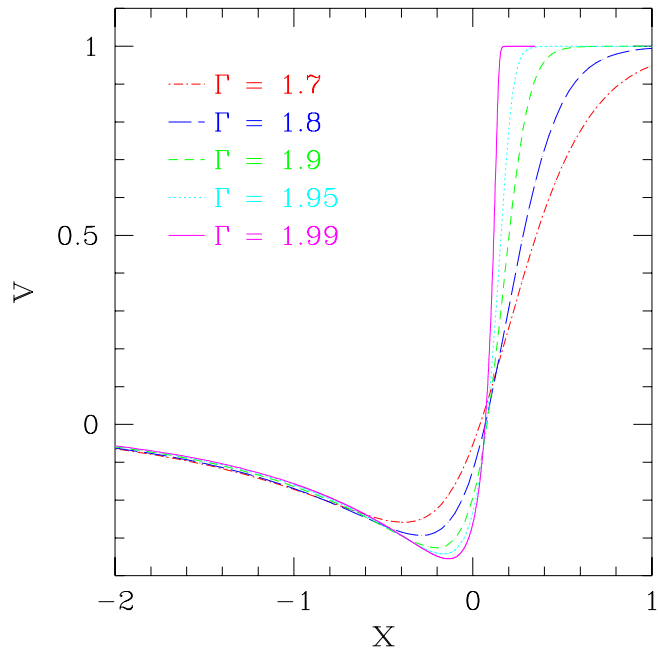}}
\caption{ The fluid velocity $v$ of the critical
solution for several values of $\Gamma$.
The sonic point is at $x=0$.
}
\label{fig:ec_v}
\end{figure}

\begin{figure}
\epsfxsize 8cm
\centerline{\epsffile{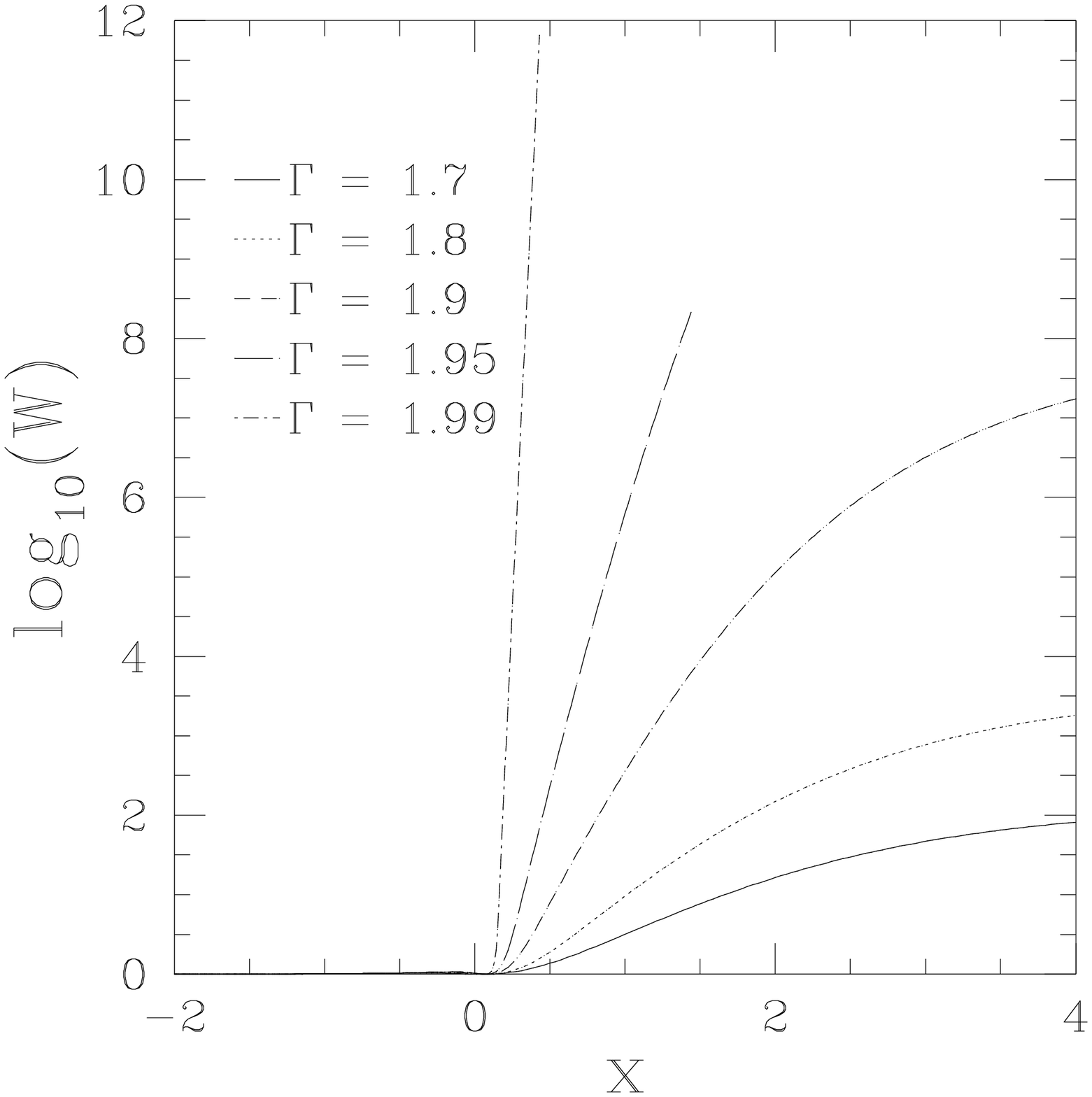}}
\caption{ The logarithm of the Lorentz factor for the fluid velocity, 
$W = 1/\sqrt{1-v^2}$, of the critical
solution for several values of $\Gamma$.
Notice the exponential growth in the Lorentz factor for larger values
of $\Gamma$.
The sonic point is at $x=0$.
}
\label{fig:ec_w}
\end{figure}

\subsection{The self-similarity {\em ansatz}}
\label{sec:ssa}

We write the spherically symmetric line element in polar-areal
coordinates as
\be
\rmd s\sq = -\alpha(r,t)\sq\,\rmd t\sq + a(r,t)\sq\,\rmd r\sq
        + r\sq\,\left( \rmd\theta\sq
        + \sin\sq\theta\,\rmd\phi\sq \right),
\label{eq:metric}
\ee
where the radial coordinate, $r$, directly measures the proper surface
area.  The Einstein equations give three relations for the metric
quantities $\alpha$ and $a$ (I:16)--(I:18) (equation numbers
with a leading `I,' refer to equations in~\cite{nc1}),
and there are two fluid equations of motion~(I:23).
A  continuously self-similar spacetime is generated by
a homothetic Killing vector $\xi$~\cite{cahill}, 
\be
\hbox{\pound}_\xi g_{ab} = 2g_{ab}.
\ee
We impose the homothetic condition and introduce coordinates $s$ and
$x$ adapted to this symmetry
\be
s \equiv  -\ln(-t), \qquad\qquad
x \equiv  \ln(-\frac{r}{t}).
\ee
The time coordinate $t$ is chosen such that the self-similar
solution reaches the origin at $t=0$, and the sonic point is at $x=0$.
We also define the dimensionless quantities
\be
N \equiv\frac{\alpha}{a\rme^x},\qquad\qquad A \equiv a^2,\qquad\qquad
\omega \equiv 4\pi r^2 a^2\rho.
\ee
The self-similarity {\em ansatz} requires that all derivatives with respect
to $s$ vanish, thus reducing the system
to autonomous first order ODEs~\cite{hka}.  These equations are presented
explicitly in \ref{sec:odeappend}; here, we formally write them as
\be
\M(y)\, y' = f(y),
\label{eq:formodes}
\ee
where $y$ is a ``state vector'' containing the four dependent variables
$y^T = (A, N, v, \omega\,)$,
$\M$ is a $4\times 4$ coefficient matrix, $f$ is a ``vector'', 
and  a prime $(')$ denotes differentiation with respect to $x$.
By construction, all solutions of the ODEs 
\eref{eq:formodes} will be continuously self-similar; to 
find the unique critical solutions, we seek CSS solutions that
are regular at both the origin and the sonic point \cite{cejc}.

\subsection{The sonic point}
\label{sec:sonicpoint}

The system of ODEs \eref{eq:formodes}
can be solved provided that the inverse matrix $\M^{-1}$ exists, 
that is, provided that the determinant~\cite{hka}
\be
\det \M \propto  \frac{(1 + N v)\sq - (\Gamma - 1)(N + v)\sq}{1 - v^2}.
\label{eq:det}
\ee
is non-zero.
When $\det \M = 0$---a condition occurring at a sonic Cauchy horizon or 
sonic point---the ODEs cannot be integrated without further assumptions.
In particular, if $\det \M = 0$, the derivatives, $y'$, may either 
(i) not exist or (ii) be undefined.
In the former case, the functions may be continuous but not differentiable,
or a shock may form (discontinuous functions) at the sonic point;
the latter case corresponds to the physically-relevant regular solutions 
in which we are interested.
By definition, the sonic point
is the position where the magnitude of the fluid velocity,
as measured by an observer
at constant $x = \ln(-r/T)$, is equal to the fluid sound speed
\be
c_s = \sqrt{\Gamma -1}. 
\ee

All CSS perfect fluid solutions that are regular at the origin
have at least one sonic point \cite{aotp}.
If $y$ is regular at the sonic point,
then the rows of $\M$ must be linearly dependent so that $\det\M = 0$.
This condition allows one to parameterize the CSS solutions with a 
{\it single} parameter, $v\sp$, which is the fluid velocity at the sonic point. 
Furthermore, this regularity condition fixes $y'\sp$ 
in terms of $v\sp$ and $v'\sp$,
yielding a quadratic condition on $v'\sp$ which we schematically write as
\be
c_2 v'_{\rm sp}{}^2 + c_1 v'\sp + c_0 = 0.
\label{eq:vprime}
\ee
Here, the coefficients, $c_0$, $c_1$ and $c_2$, 
are complicated functions of $y\sp$, and for
simplicity of presentation will not be given explicitly.  
The key point is that this
constraint---that the critical solutions are regular at the sonic 
point---limits the number of solutions to discrete
values of $v\sp$,  and virtually eliminates the possibility 
that globally regular solutions
with more than one sonic point exist \cite{aotp,tfrh}.
Indeed, all of the $\Gamma \le 2$
critical solutions we have found, either from a CSS {\em ansatz,} or 
by solving the full Einstein/fluid equations, 
have only one sonic point.

\subsection{Solving the ODEs}
\label{sec:solveode}

The system of ODEs \eref{eq:formodes} is solved 
by choosing a candidate fluid velocity, $v\sp$, at the sonic
point, and integrating numerically from the sonic point
toward the origin. 
The inward integration is halted when either $A < 1$ or $\det \M = 0$, 
and these generic stopping criteria allow one to determine the
parameter $v\sp$ by a standard ``shooting'' procedure.  (If
$A < 1$ signals that $v\sp$  is too small, then $\det \M = 0$ 
indicates that $v\sp$ is too large, and vice versa).
Once $v\sp$ has been determined from the inward integration,
the solution can be completed by
integrating outwards from the sonic point.   
The entire solution process is complicated by the fact that 
the integration can not actually begin at the sonic point, since
$\det \M=0$ there.  Therefore, we first expand 
the dependent variables $y$ about the sonic point to first order
\be
y_o \approx y\sp + y'\sp \triangle x,
\ee
where $\triangle x \equiv x_o -x\sp$, and actually begin the integration 
from $x_o$.
$\triangle x$ is chosen so that the $\Or((\triangle x)^2)$ error terms 
in the expansion,
are smaller than the error tolerance
allowed in the solution.  We obtain $y'\sp$ by solving
\eref{eq:vprime} for $v'\sp$ and integrate 
the ODEs for both roots.  

The ODEs are integrated using \LSODE, a robust numerical
routine for integrating ODEs \cite{lsode1,lsode2}, 
and {\it all} of the critical solutions can be found using 
double precision arithmetic, 
{\it except} those solutions for $\Gamma \approx 1.89$.  
These $\Gamma\approx 1.89$ solutions require greater precision,
and can be found using 
the arbitrary precision implementation of \LSODE\ in {\sl Maple V,}
which also proved invaluable for convergence testing the solutions.
In the convergence tests we vary
$\triangle x$, the \LSODE\ absolute error tolerance $\varepsilon$
(the relative error tolerance is set to zero), 
and the number of digits used in the calculation,
while monitoring the residual of the algebraic constraint (\ref{eq:resid})
as an indication of the error in the solution.
For example, we calculated the critical solution for $\Gamma=1.99$
using 40 digits and error tolerances $\varepsilon = 10^{-10}$, 
$10^{-15}$, $10^{-20}$
and $10^{-25}$, and then performed similar tests using 30 and 35 digits.
In all cases the solutions converge, and the
residual of \eref{eq:resid} is $\Or(\varepsilon)$.

The critical solutions for
several values of $\Gamma$ are shown in figures
\ref{fig:ec_a}--\ref{fig:ec_w}. 
One notes that as $\Gamma \rightarrow 2$, 
it becomes increasingly difficult to integrate
outwards from the sonic point, and this limitation is especially apparent in
the Lorentz factors for the $\Gamma = 1.95$ and $\Gamma =1.99$ 
critical solutions shown in \fref{fig:ec_w}.
The ODE solver fails when the Jacobian matrix,
\be
{\cal J} = \frac{\PD \lp \M^{-1}f \rp}{\PD y},
\ee
becomes ill-conditioned,
which often occurs when $\omega \approx \varepsilon$ and/or
$v \approx 1 - \varepsilon$, where $\varepsilon$ is the error 
tolerance supplied to \LSODE.
We can extend these solutions further outwards simply by increasing the 
number of digits used in the calculation.  However the increase
in maximum physical radius as a function of resolution is very 
modest---since the Lorentz factor $W$ increases exponentially with
increasing radius---and the process quickly becomes prohibitively expensive.
Clearly, $\cal J$ is always poorly conditioned near sonic points, but 
$\cal J$ is particularly 
poorly conditioned near the sonic point for $\Gamma \approx 1.89$---one 
can view this as a clear signal that numerical work 
will be difficult in the $\Gamma \approx 1.89$ regime.

\subsection{Nature of the sonic point for $\Gamma \gtrsim 1.89$}
\label{sec:sonicpoint2}

\begin{figure}
\epsfxsize 7.8cm
\centerline{\epsffile{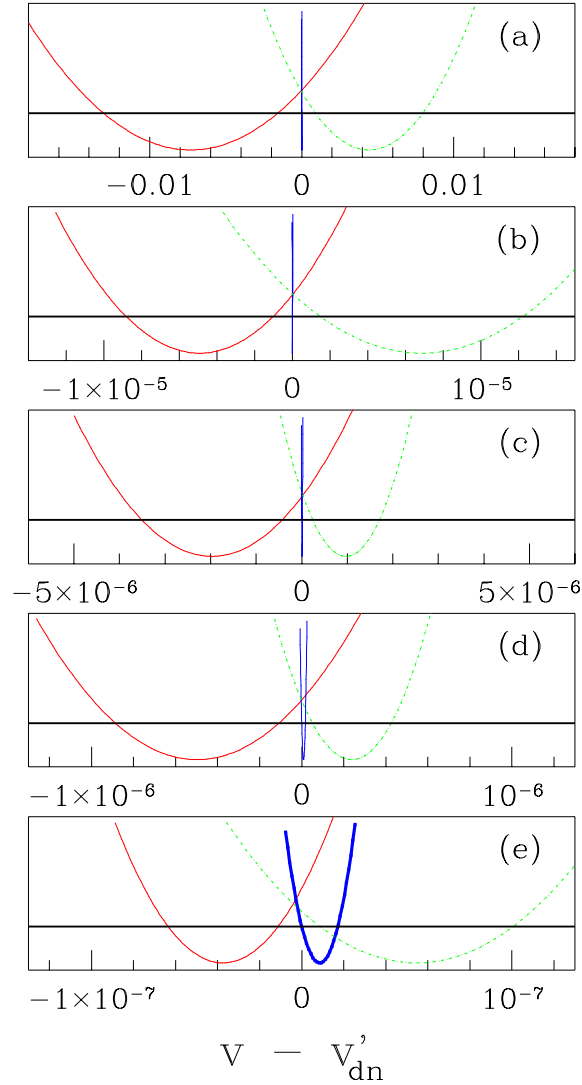}}
\caption{
The sonic point of the $\Gamma\dn\simeq 1.8896244$ critical solution
is a degenerate node,
and this figure shows the approach to degeneracy
by plotting the quadratic function in
\eref{eq:vprime} as $\Gamma \rightarrow \Gamma\dn$.
The roots of \eref{eq:vprime} are possible values for $v'$ at the
sonic point, and when $\Gamma = \Gamma\dn$ these roots are equal, $v'\dn$.
Each frame shows a parabola for three values of $\Gamma$.  The
parabolas on the left are for $\Gamma$ less than $\Gamma\dn$
($\Gamma_<$), and
the parabolas on the right are for $\Gamma$ greater than
$\Gamma\dn$ ($\Gamma_>$).  The center parabola has the same 
$\Gamma$ in all five frames, namely $\Gamma_C\simeq \Gamma\dn$
($\Gamma_C = 1.88962441796875$),
and one must be careful not to mistake it for a vertical axis
in frames (a)--(c).  For all $\Gamma$,
the critical solution's $v'\sp$ 
is the root closest to the center ($v'\dn$).
We estimate $\Gamma\dn \simeq 1.8896244169921874$ ($v\sp = -0.18696\ldots\,$,
$v'\sp = 1.7385\ldots\,$,
and $\vartheta\sp =1.0000000002$), and for clarity, we use  $v'\dn$ to 
translate the horizontal axis such that the parabolas cluster 
around zero, and normalize the parabolas.
In frame~(a), 
        $\Gamma_<=1.889$                 and $\Gamma_>=1.890$.
In (b), $\Gamma_<=1.889624$              and $\Gamma_>=1.889625$.
In (c), $\Gamma_<=1.88962425$            and $\Gamma_>=1.8896245$.
In (d), $\Gamma_<=1.889624375$           and $\Gamma_>=1.8896244375$.
In (e), $\Gamma_<=1.8896244140625$       and $\Gamma_>=1.889624421875$.
These calculations were done with {\sl Maple V} using 30 digits and
$\varepsilon = 10^{-18}$.
}
\label{fig:fvp}
\end{figure}

Although nonlinear systems of ODEs  are often impossible to solve in 
closed-form, qualitative features of their solutions can frequently be 
deduced by linearizing the equations about ``critical'' points.
Perfect fluid CSS solutions have often been studied
using this type of analysis
\cite{dm, hka, goliath, aotp, ob, bh1, bh2, tfrh},
and here we discuss some of these results in the context of our work. 
We emphasize, however, that we have {\em not} performed perturbation 
analyses in our current work.

The CSS perfect fluid equations \eref{eq:formodes} can be linearized
about the sonic point, resulting in a system we can write in the 
form
\be
y'=\B\,y,
\label{eq:lineq}
\ee
where $\B$ is a matrix which, generically, has 
two non-zero eigenvalues, which we label $\kappa_1$ and $\kappa_2$
(or simply $\kappa$ when the distinction is irrelevant),
with corresponding eigenvectors $V_1$ and $V_2$.
Near the sonic point, the solution of the linear equations \eref{eq:lineq}
can be written~\cite{bh1}
\be
y = y\sp + k_1V_1\e^{\kappa_1 x} +  k_2V_2\e^{\kappa_2 x},
\ee
where $k_1$ and $k_2$ are arbitrary constants.  The eigenvalues $\kappa$
provide important information about the solutions near the sonic point,
and we classify the sonic point according to the relative values of
$\kappa$, as given by the quantity $\vartheta$ \cite{aotp}
\be
\vartheta = \frac{\Gamma + \sqrt{\eta}}{\Gamma - \sqrt{\eta}}.
\ee
Here, 
\beq
&\eta \equiv 4(3\Gamma - 2)U^2 - (3\Gamma^2 - 12\Gamma + 8)(1-4U),
\eeq
and
\beq
&U \equiv \frac{A - 1}{2\omega}.
\eeq
The sonic point classification in terms of $\vartheta$ is shown in
\tref{table2}.  Due to the facts that (i) $\vartheta$ is only a function 
of $y\sp$, 
and (ii) the eigenvalues $\kappa$ are related to $v'\sp$ \cite{tfrh},
we can make a connection to the linearized theory without explicit calculation
of the eigenvalues.

\Table{Classifications of the sonic point using $\vartheta$.
\label{table2}}
\br
$\vartheta < 0$& saddle point\\
$\vartheta > 0$& nodal point\\
$\vartheta \in \mathbb C$ & focal point\\
$\vartheta=1$& degenerate nodal point\\
\br
\endTable

Maison \cite{dm} and Goliath \etal \cite{goliath} have previously concluded
that
the sonic points for $\Gamma \gtrsim 1.89$ are foci, with complex $\kappa$
and $v'\sp$, and hence have suggested that physical self-similar solutions
do not exist for $\Gamma\gtrsim 1.89$. 
(Hara \etal did not address the existence of 
solutions for $\Gamma\gtrsim 1.89$, but presumably also encountered problems 
with their numerical analysis in that regime.)
However, we find that $\vartheta > 0$ for {\em all}
$\Gamma \gtrsim 1.89$ critical solutions, and thus conclude that
the sonic-points for those solutions are nodes rather than foci.
It seems plausible that this apparent contradiction stems from 
insufficient numerical precision in the earlier studies.
To provide some specific evidence to back this claim, we have used 
{\sl Maple} with 30 digits
to find a critical solution for $\Gamma\simeq\Gamma\dn$.
Then, taking $v\sp$ from this solution, we have calculated $\vartheta$
using both 30 digits in {\sl Maple} and {\tt FORTRAN} double precision.
The {\tt FORTRAN} calculation gave a complex $\vartheta$---which 
would support the erroneous (we claim) conclusion that the sonic point is a 
focus. The same calculation done with greater precision using {\sl Maple}
shows that the sonic point is actually a node.
In addition, we find that for
$\Gamma < \Gamma\dn$ (with $\Gamma$ restricted to $\Gamma > 1.8$ 
for simplicity, and $\Gamma\dn$ defined below),
the critical solution's $v'\sp$ is the maximum root of \eref{eq:vprime}, 
while for $\Gamma > \Gamma\dn$, the critical solution's $v'\sp$ is
the minimum root.
As $\Gamma\rarrow\Gamma\dn$, the two roots $v'\sp$ come closer together
until they are equal for $\Gamma\dn$, as shown in \fref{fig:fvp}. 
Here the sonic point is
a degenerate node with $\Gamma\dn \simeq 1.8896244$
($\eta\dn = \Or(\varepsilon)$, and 
$\vartheta\dn = 1 + \Or(\sqrt{\varepsilon})$).


\section{Simulation of critical solutions}
\label{sec:results}

A crucial check that the CSS solutions of the ODEs are indeed
the unique critical solutions involves a comparison
with the solutions of the full Einstein/fluid equations.
Although relativistic fluid equations are known to be difficult to solve
(particularly relative to ``fundamental-field'' equations, such as the 
EMKG system), modern high-resolution shock-capturing 
schemes~\cite{rjlbook,rjl98,ibanez,romero} allow
one to calculate flows with shocks of almost arbitrary strength.
However, perhaps the greatest challenge in finding the perfect fluid critical 
solutions---especially as $\Gamma \rarrow 2$---is the accurate treatment of 
flows with very large Lorentz factors.
In \cite{nc1} we have outlined our computer program which solves
the spherically-symmetric Einstein/fluid system, and using this 
code we have found
the perfect fluid critical solutions for $1.05 \le \Gamma \le 2$.
All of these solutions
are continuously self-similar (CSS) and black hole formation for
near-critical initial data begins with
infinitesimal mass (Type~II).
In this section we compare
the ODE and PDE solutions, discuss the mass-scaling
exponents $\gamma$, and finally, briefly discuss critical solutions
for the ideal-gas equation of state.

\begin{figure}
\epsfxsize = 10cm
\centerline{\epsffile{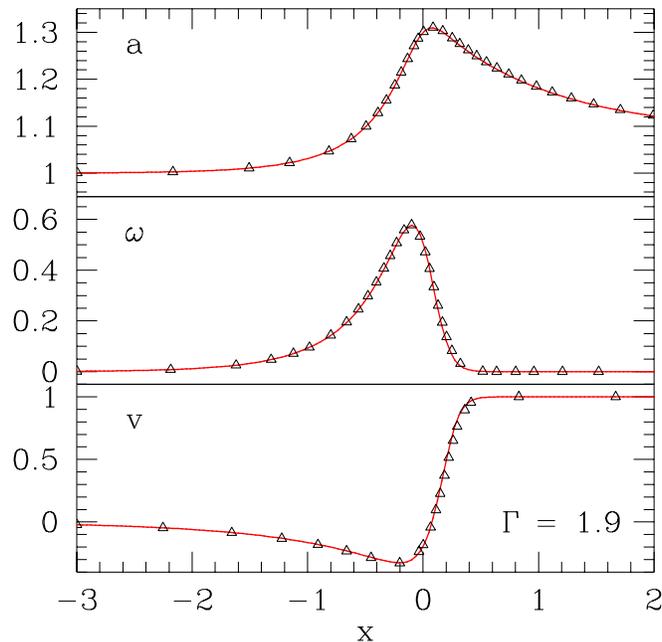}}
\caption{The $\Gamma = 1.9$ critical solution.
The solid lines are the solution obtained by solving the ODEs,
and the triangles indicate selected points from the
solution of the PDEs.
}
\label{fig:g19}
\end{figure}

\begin{figure}
\epsfxsize = 10cm
\centerline{\epsffile{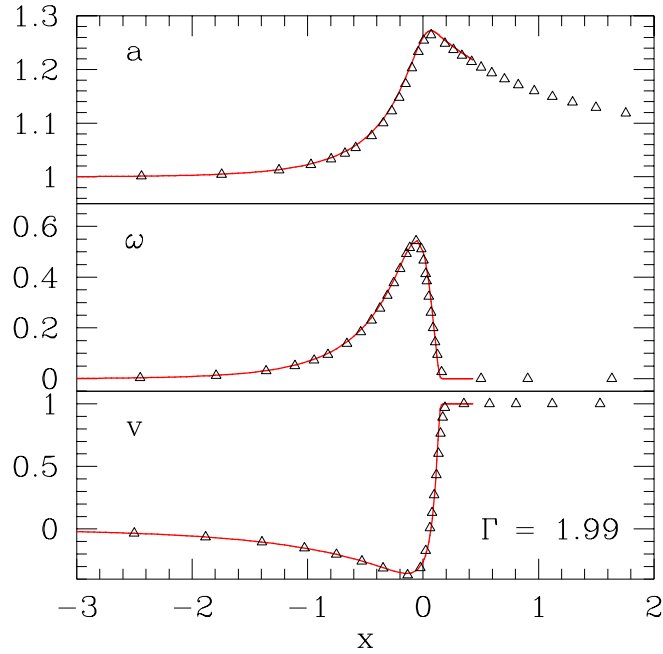}}
\caption{The $\Gamma = 1.99$ critical solution.
The solid lines are the solution obtained by solving the ODEs,
and the triangles are selected
points from the PDE solution.
The PDE solution underestimates
the fluid density in the pulse leading, to a corresponding error in $a$.
This problem stems from a lack of resolution in the computational
grid.
}
\label{fig:g199}
\end{figure}

\subsection{Critical solutions}
\label{sec:sols}

Figures \ref{fig:g19}--\ref{fig:g199} show the PDE 
critical solutions for $\Gamma = 1.9$ and $\Gamma = 1.99$ respectively,
and include the ODE solutions for comparison.
The $\Gamma=1.9$ PDE and ODE
solutions compare well,
while \fref{fig:g199} indicates that the $\Gamma = 1.99$ PDE solution
underestimates $a$ and $\omega$.
From our experience with other critical solution searches,
we feel that this discrepancy is the result of insufficient 
(spatial) resolution. 
As $\Gamma\rightarrow 2$, the fluid becomes increasingly dynamic,
requiring greater precision to resolve the solution's relevant
features, and it becomes increasingly expensive
to calculate the critical solutions. 
The mass-scaling exponents $\gamma$ shown in \tref{table1}
provide a quick guide to the requisite dynamical range for 
a critical evolution
as $\Gamma\rarrow 2$.
As $(p-p^\star)$ changes by $n$ orders of magnitude, the relevant
length scales in the solution, 
such as the radius of a black hole $R_{\rm BH}$,
change by $\gamma n$ orders of magnitude.  
The mass-scaling exponent for the stiff
fluid, $\gamma \approx 1$, is almost three times larger
than the scaling exponents for a radiation fluid ($\gamma \approx 0.36$)
or massless scalar field ($\gamma_{\rm SF} \approx 0.37$),
and simulations of the stiff fluid critical
solutions require correspondingly more resolution.

\begin{figure}
\epsfxsize = 10cm
\centerline{\epsffile{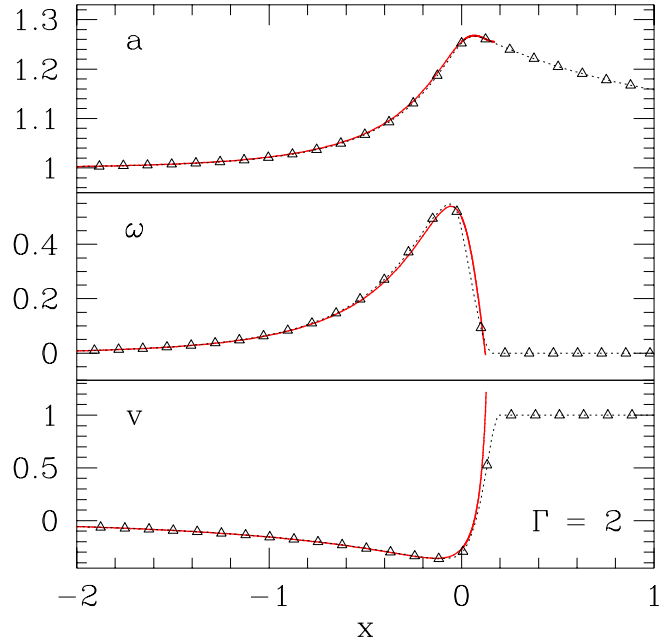}}
\caption{The $\Gamma=2$ critical solution.  The solid lines show
the solution obtained from the ODEs, and the dotted lines with 
triangles show the solution obtained by solving the PDEs.
Here some divergence between the PDE and ODE solutions can be
seen.  The PDE solution for $\omega$ lies {\it above} the ODE solution 
for $x <0$, behavior opposite from that observed in the $\Gamma=1.99$
solution.  Beyond the sonic point, $x=0$, the solutions become
very different.  Here the ODE solution becomes unphysical---in the
traditional understanding of a perfect fluid---with
$\omega < 0$ and $v>1$.
The PDE solution retains physical values for $v$
and $\omega$---$v < 1$ and $\omega > 0$---because these values are 
constrained to the physical regime in the evolution code~\cite{nc1}.}
\label{fig:g2sol}
\end{figure}

The critical solutions for $\Gamma<2$ all appear very similar; indeed
one can imagine that one could smoothly transform a solution for a given
$\Gamma$ into a solution for a different $\Gamma$.  At first glance,
the $\Gamma=2$ solution seems to fit nicely into this ``family'' of
critical solutions parameterized by $\Gamma$---it is CSS, Type II,
and differs only slightly from the $\Gamma=1.99$ critical solution.
However, the ODE solution (obtained by solving ODEs with the CSS
{\it ansatz\/}) indicates that important differences may exist between
the $\Gamma=2$ and $\Gamma<2$ critical solutions.  As noted previously,
we are unable to integrate the ODEs for $\Gamma$ near 2 to arbitrarily
large $x$.  In these cases, we observe that the Lorentz factor, $W$,
grows exponentially (see \fref{fig:ec_w}), with a corresponding
exponential decrease in $\omega$, until {\tt LSODE} is unable to
satisfy the required error tolerances.  We emphasize that in these 
$\Gamma<2$ solutions, the fluid velocity and density retains its
expected ``physical'' properties: $\omega > 0$, and $|v|< 1$.
The $\Gamma=2$ solution (see \fref{fig:g2sol}), on the other
hand, displays very different behavior.  Instead of the exponential
approach of $v\to 1$ and $\omega\to 0$, we find that $\omega$ and $v$
pass through their expected physical bounds, giving $\omega < 0$ and
$v > 1$.  As is generally known, the stiff perfect fluid can be
related to a scalar field, and, very recently, a CSS scalar field 
solution has been found by Brady and Gundlach that exactly matches 
the ODE $\Gamma=2$ fluid solution~\cite{bgnc}.

The $\Gamma=2$ PDE solution is also shown in \fref{fig:g2sol},
and in contrast to the ODE solution, this solution retains 
``physical'' values for the fluid variables.  However, this property
of the PDE solution is achieved by {\it fiat:} we impose a ``floor''
on the fluid variables such that $\rho > 0$ and $|v|<1$~\cite{nc1}.
While the floor is used generally for $\Gamma\gtrsim 1.8$ without
noticeable ill effect for $\Gamma<2$, it is clear that the floor
affects the $\Gamma=2$ solution (as compared to the ODE solution)
even in the regime where $\omega>0$ and $|v|<1$.  The mathematical
and physical significance of this observation is clearly an issue 
which requires more study---for example, can a vacuum region be 
matched to the $\Gamma=2$ fluid in such a way
that the fluid remains equivalent to a EMKG field~\cite{gundlach_pc}?

Finally, we note that the $\Gamma=2$ critical solution is not
related to other familiar EMKG solutions, such as
the Roberts solution~\cite{rob1, rob2}, 
or the EMKG critical solution~\cite{mwc93}.
Both of these solutions have spacelike gradients of the scalar field.
Extracting data from a near-critical $\Gamma=2$ solution, we have
set equivalent initial data for an EMKG evolution and  then have evolved
the data with the Einstein/scalar equations of motion.  However,
the evolution of the scalar field
does {\it not} seem to match the $\Gamma=2$ perfect-fluid critical solution
for any appreciable length of time,
and na\"\i ve variations of the initial data 
apparently produce the usual DSS scalar
field critical solution at the black-hole threshold.

\subsection{Mass-scaling exponents}

\begin{figure}
\epsfxsize = 8cm
\centerline{\epsffile{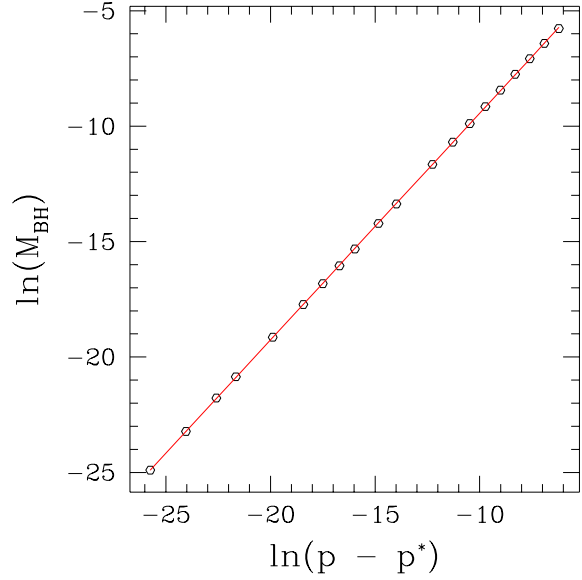}}
\caption{Illustration of black-hole mass scaling for the case
$\Gamma =2$.  In this instance---as for all values of $\Gamma$ considered
here---the critical behavior is Type II,  allowing one to create
arbitrarily small black
holes through sufficient fine-tuning of initial data.
}
\label{fig:g2bhm}
\end{figure}

Mass-scaling exponents $\gamma$ are found by evolving near-critical
initial data sets which lead to the formation of black holes. 
In our coordinate system, black hole formation is signaled by
\be
\left.\frac{2m(r,t)}{r}\right|_{R_{\rm {BH}}}\rightarrow 1,
\ee
where $R_{\rm {BH}}$ is the (areal) radius of the black hole.  The 
black hole mass is then simply given by
\be
M_{\rm {BH}} = \frac{R_{\rm {BH}}}{2}.
\ee
As mentioned earlier, all of the critical solutions discussed here 
are Type~II,
meaning that the associated black-hole transition begins with 
infinitesimal mass.
As a typical example of our results, the mass-scaling of 
near-critical solutions for $\Gamma = 2$ is shown in \fref{fig:g2bhm}.

The simple adaptive grid that we use~\cite{nc1}
did not allow us to calculate
$M_{\rm {BH}}$ with sufficient accuracy to justify searching for
$p^\star$ to the limit of machine precision in a reasonable
amount of time.  We therefore
estimated $p^\star$ by searching for the best linear fit to
\be
\ln M_{\rm {BH}} \propto \gamma \ln\left|p-p^\star\right|.
\ee
The totality of mass-scaling exponents $\gamma$
calculated from our simulation data 
are shown in \tref{table1}, along with the values predicted 
from Maison's perturbative calculations~\cite{dm}.
(These exponents for $\Gamma \le 1.98$ are similar to those
found independently by Brady and Cai~\cite{brady_cai}.)
For a variety of reasons, estimation of the error (no doubt 
overwhelmingly ``systematic'') in the mass-scaling 
exponents is not an easy task, and we therefore have provided 
estimates of $\gamma$ which are conservative in their use 
of ``significant'' digits. 
One notes that as $\Gamma \rarrow 2$,
the mass-scaling exponent $\gamma \rarrow 1$.  This trend suggests
that $\gamma = 1$ for $\Gamma=2$.  However our numerical results can
only determine $\gamma$ to very limited precision, and we are not
currently aware of any argument for a precise equality.

\Table{The mass-scaling exponent $\gamma$ as
a function of the adiabatic constant $\Gamma$.  The second column
shows the mass-scaling exponents estimated from $M_{\rm {BH}}(p)$ by
evolving
near-critical ($p\to p^\star$) initial data. 
For comparison, 
``calculated'' exponents---computed from
perturbative calculations---are also listed.\label{table1}
}
\br
&&\centre{2}{$\gamma$}\\
\ns
&&\crule{2}\\
$\Gamma$&&Measured & Calculated$^{\rm a}$\\
\mr
\lineup
1.05  && 0.15 & 0.1478 \\
1.1   && 0.19 & 0.1875 \\
1.2   && 0.26 & 0.2614 \\
1.3   && 0.33 & 0.3322 \\
4/3   &$\quad$& 0.36 & 0.3558 \\
1.4   && 0.40 & 0.4035 \\
1.5   && 0.48 & 0.4774 \\
1.6   && 0.56 & 0.5556 \\
1.7   && 0.64 & 0.6392 \\
1.8   && 0.73 &0.7294  \\
1.888 && 0.82 & 0.8157 \\
1.89  && 0.82 & ---\\
1.9   && 0.83 & ---\\
1.92  && 0.86 & ---\\
1.95  && 0.9  & ---\\
1.99  && 1    & ---\\
2     && 1    & ---\\
\br
\end{tabular}
\item[] $^{\rm a}$ From Maison \cite{dm}.
\end{indented}
\end{table}

\subsection{Critical solutions for the ideal gas}

\begin{figure}
\epsfxsize = 9cm
\centerline{\epsffile{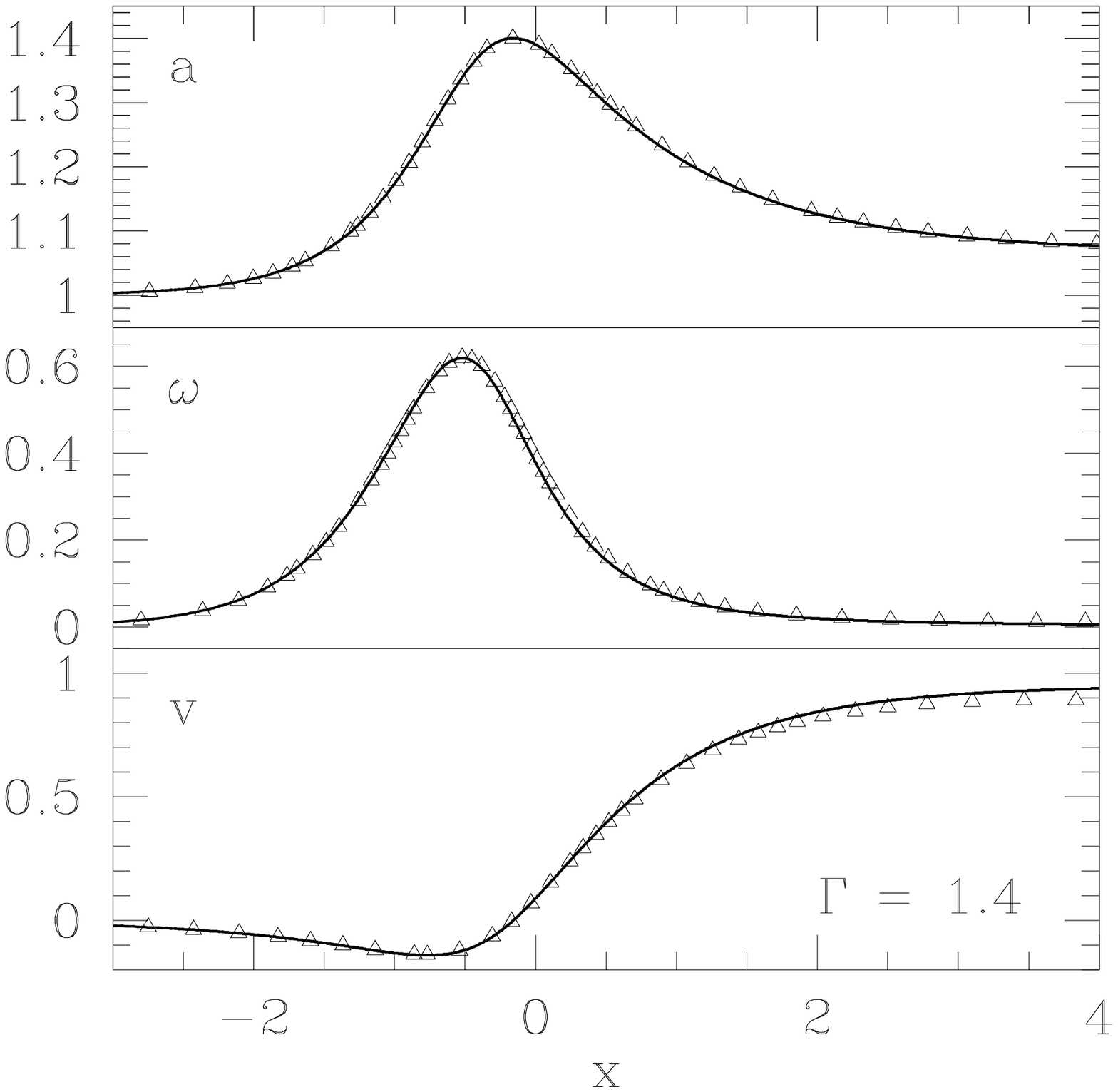}}
\caption{A comparison of the critical solutions for the
ideal-gas and ultrarelativistic equations of state for $\Gamma = 1.4$.
The solid lines are solutions obtained by solving the ODEs with the
ultrarelativistic EOS,
and the triangles are selected
points from the PDE solution which uses the ideal-gas EOS.  
The ideal gas is in the ultrarelativistic limit near the
infalling matter (see \fref{fig:p-rho-log}), and the 
two solutions correspond in this region.  At large $r$
the ultrarelativistic approximation breaks down, and the solutions
differ.
The ideal-gas EOS
solution was computed using a code similar to the one 
described in~\cite{nc1}. 
}
\label{fig:ad-er}
\end{figure}

\begin{figure}
\epsfxsize = 8cm
\centerline{\epsffile{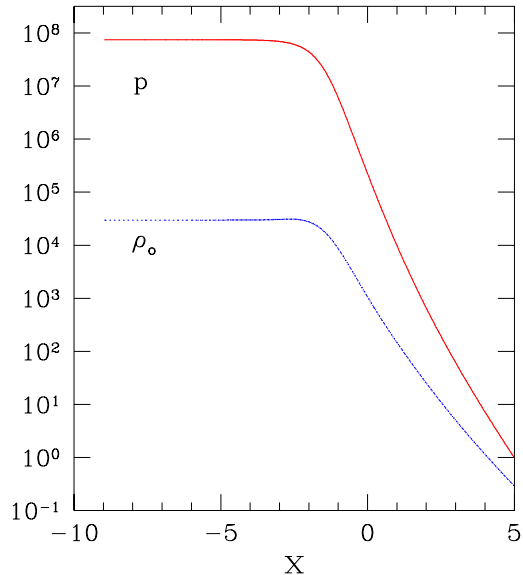}}
\caption{$P$ and $\rho_o$ are shown at the moment of maximum
compression in this log-log plot of a marginally subcritical evolution of
a $\Gamma = 1.4$ ideal gas.  Note that near the origin, the ideal gas
is clearly in 
the ultrarelativistic limit where
$P\gg \rho_o$.  At large $r$ ($X$), the ultrarelativistic
limit no longer holds, and the solution does not match
one computed using the ultrarelativistic EOS.
}
\label{fig:p-rho-log}
\end{figure}

The equation of state  (EOS) $P=(\Gamma -1)\rho$ can
be interpreted as the ultrarelativistic limit of the ideal-gas
EOS 
\be
P=(\Gamma -1)\rho_o\epsilon,
\ee
where $\rho_o$ is the rest energy density
and $\epsilon$ is the specific internal energy density.
Following Ori and Piran~\cite{aotp},
Evans~\cite{ce93} (and others), 
we have argued~\cite{nc1} that self-similar perfect 
fluid solutions {\em require}
the ultrarelativistic EOS.  Let us now consider searching for
critical solutions with the ideal-gas EOS.
Heuristically, one can describe critical behavior in terms of competition
between the fluid's kinetic energy and gravitational potential energy.
One might expect that in the critical solution, which stands just on the
verge of black-hole formation, $P \gg \rho_o$, and that the
 critical solutions for the ideal-gas EOS
would correspond to the ultrarelativistic EOS
solutions. Using an evolution code, we have found critical solutions
for the ideal-gas EOS, and these solutions
{\em do} match the corresponding solutions with the
ultrarelativistic EOS.  
As an example,
the critical solution for a $\Gamma =1.4$ ideal gas
is compared with the precisely CSS ultrarelativistic
$\Gamma=1.4$ solution in \fref{fig:ad-er}.
Additional evidence that near-critical ideal gas 
solutions are ultrarelativistic is shown in 
\fref{fig:p-rho-log}.

\section{Conclusion}
\label{sec:conclusion}

Following seminal work by Evans and Coleman~\cite{cejc}, 
Maison~\cite{dm}, and Hara, Koike and Adachi~\cite{hka} 
have previously reported the existence
of CSS solutions with a single unstable (relevant) mode for 
$\Gamma \lesssim 1.89$.
In this paper we have shown that such solutions 
also exist for $\Gamma \gtrsim 1.89$, and have 
found evidence that the sonic point is a degenerate {\em node} for the 
$\Gamma\dn\simeq 1.8896244$
critical solution.  Our results come from (i) a demonstration
of the existence of globally regular CSS solutions for $\Gamma\le 2$ 
(starting from a CSS {\em ansatz\/}),  and 
(ii) the evolution of finely-tuned initial data in
full simulations of the Einstein/fluid equations.  The simulations verify 
that the CSS solutions sit at the threshold of black hole formation, and
also allow us to compute mass scaling exponents which are in good 
agreement with predictions from perturbation theory.
The $\Gamma=2$ critical solution is also CSS and Type~II, with 
a mass-scaling exponent very close to (if not identical to) unity.
We have also investigated critical collapse using the ideal-gas
equation of state, and, as expected, have found that
the fluid {\em is} well-described in the near-critical regime 
by an ultrarelativistic approximation, 
and thus has critical solutions identical to those generated using 
$P = (\Gamma - 1) \rho$.

While we have addressed some of the questions regarding 
the $\Gamma\gtrsim 1.89$ critical solutions
that have appeared in the literature, other avenues for further research
remain.  Perhaps foremost is the need for a greater understanding
of the $\Gamma=2$ critical solution, both in a more precise estimate
of its mass-scaling exponent, and its precise relation to the
EMKG field.  
In addition, Gundlach~\cite{gund1} and Martin-Garcia and Gundlach~\cite{gund2}
have recently investigated the critical behavior of 
an axisymmetric radiation fluid with angular momentum.
Full simulations of such scenarios will undoubtedly yield 
further interesting results.

Finally, we wish to point out an obvious use  of critical solutions 
for testing and benchmarking new codes which solve the 
equations of motion (PDEs) for general relativistic fluids.  
In particular, we are unaware
of widely-used test-bed solutions which combine
both dynamic fluid flows {\it and} strong gravitational fields.
Common tests for such codes currently include
Riemann shock tube configurations, static Tolman-Oppenheimer-Volkoff
solutions and Oppenheimer-Synder dust collapse.  
The former tests lack gravitational effects, and the latter tests 
lack either dynamic fluid flows or pressure effects.
While we do not advocate abandoning traditional 
tests, we wish to emphasize that critical solutions are 
novel in that they combine several 
computationally challenging characteristics:
extremely relativistic and dynamic fluid flows, 
strong gravitational field dynamics, 
and relevant features over many space/time scales.
In particular, as adaptive mesh refinement (AMR) becomes available 
for multi-dimensional simulations, these features should make critical 
solutions ideal candidates for code calibration.

\section*{Acknowledgments}
\label{sec:ack}

This work was supported in part by the National Science
Foundation under Grants PHY93-18152 (ARPA supplemented), PHY94-07194, 
PHY97-22068, by a Texas
Advanced Research Project grant, and by an NPACI award of computer time.
We thank P.~Brady, D.~Garfinkle, M.~Goliath, C.~Gundlach, and D.~Maison for 
helpful discussions and suggestions, and are particularly 
indebted to M.R.~Dubal for his substantial contributions 
to an earlier (and unpublished) attack on the problem of 
perfect-fluid critical collapse.
MWC gratefully acknowledges the 
hospitality of the Institute for Theoretical Physics, UC Santa Barbara, 
where part of this research was carried out.

\appendix
\section{ODEs for a self-similar spacetime}
\label{sec:odeappend}

The Einstein equations for a spherically symmetric, CSS perfect fluid
are presented in this appendix for reference.
These equations are derived in \cite{hka}.

\be
\frac{A'}{A} = 1 - A + \frac{2\omega\lp 1 + \lp\Gamma - 1\rp v^2\rp}{1-v^2}
\label{eq:a1}
\ee
\be
\frac{N'}{N} = -2 + A - \lp 2 - \Gamma \rp\omega
\label{eq:a2}
\ee
\be
\frac{A'}{A} = -\frac{2\Gamma N v \omega}{1-v^2}
\label{eq:a3}
\ee
\beq
&\lp 1 + N v\rp\frac{\omega'}{\omega} + \frac{\Gamma\lp N + v\rp v'}{1-v^2}
\nonumber\\
&\qquad = \frac{3}{2}(2-\Gamma)N v - \frac{2+\Gamma}{2}A N v 
        + (2-\Gamma)N v\omega
\label{eq:a4}
\eeq
\beq
&(\Gamma - 1)(N + v)\frac{\omega'}{\omega}
+ \frac{\Gamma(1 + N v)v'}{1-v^2}\nonumber\\
&\qquad = (2-\Gamma)(\Gamma - 1)N\omega + \frac{7\Gamma -6}{2}N 
+ \frac{2-3\Gamma}{2}AN
\label{eq:a5}
\eeq
The equations (\ref{eq:a2})--(\ref{eq:a5}) are integrated to find the critical
solutions, and comprise the system $\M\, y'=f$ in \eref{eq:formodes}.
Equations (\ref{eq:a1}) and (\ref{eq:a3}) can be combined to give
an algebraic constraint
\be
(1 - A)(1 - v^2) + 2\omega\lp 1 + (\Gamma - 1)v^2\rp  +2\Gamma N v\omega = 0.
\label{eq:resid}
\ee
This equation is {\it not} used to integrate the ODEs, but
is monitored during the 
integration as an indication of the error in the solution.

\section*{References}
\begin{thebibliography}{99}

\bibitem{mwc93}   Choptuik M W 1993  \PRL {\bf 70} 9

\bibitem{cg97}    Gundlach C 1998 {\it Adv.\ Theor.\ Math.\ Phys.} {\bf 2} 1;
                  {\tt gr-qc/9712084}

\bibitem{mwc98}   Choptuik M W 1998 {\tt gr-qc/9803075}

\bibitem{dm}      Maison D 1996 \PL B {\bf 366} 82

\bibitem{hka}     Hara T, Koike T, and Adachi S 1996 {\tt gr-qc/9607010}

\bibitem{kha2}    Koike T, Hara T, and Adachi S 1999 \PR D {\bf 59} 

\bibitem{kha95}   Koike T, Hara T and Adachi S 1995 \PRL {\bf 74} 5170

\bibitem{mwc_ym}  Choptuik M W, Chmaj T and Bizon P 1996 \PRL
                     {\bf 77} 424

\bibitem{pb_msf}  Brady P 1997 {\tt gr-qc/9709014}

\bibitem{cc1}     Carr B J and Coley A A 1998 {\tt gr-qc/9806048}

\bibitem{goliath} Goliath M, Nilsson U S and Uggla C 1998 \CQG
                      {\bf 15} 2841

\bibitem{cc2}     Carr B J and Coley A A 1999 {\tt gr-qc/9901050}

\bibitem{cejc}    Evans C R and Coleman J S 1994 \PRL {\bf 72} 1782

\bibitem{perkins} Perkins T J W 1996 thesis (unpublished)

\bibitem{carr1}   Carr B J, Coley A A, Goliath M, Nilsson U S and Uggla C
                  1999 {\tt gr-qc/990131}

\bibitem{carr2}   Carr B J, Coley A A, Goliath M, Nilsson U S and Uggla C
                  1999 {\tt gr-qc/990270}

\bibitem{brady_cai} Brady P R and Cai M J 1998 {\tt gr-qc/9812071}

\bibitem{taub}    Taub A H 1959 {\it Arch.\ Rat.\ Mech.\ Anal.} {\bf 3} 312

\bibitem{mad1}    Madsen M K 1985 {\it Astrophys. Space Sci.}
                          {\bf 113} 205

\bibitem{mad2}    Madsen M K 1988 \CQG {\bf 5} 627

\bibitem{nc1}     Neilsen D W and Choptuik M W 1998 submitted to \CQG

\bibitem{aotp}    Ori A and Piran T 1990 \PR D {\bf 42} 1068

\bibitem{ce93}    Evans C R 1993 in
                     {\it Lecture Notes of the Numerical Relativity
                     Conference, Penn State University, 1993} (unpublished)

\bibitem{ob}      Bogoyavlenskii O I 1977 {\it Sov.\ Phys.\ JETP}
                            {\bf 46} 633

\bibitem{bh1}     Bicknell G V and Henriksen R N 1978
                         {\it ApJ} {\bf 219} 1043

\bibitem{bh2}     Bicknell G V and Henriksen R N 1978
                         {\it ApJ} {\bf 225} 237

\bibitem{tfrh}    Foglizzo T and Henriksen R N 1993 \PR D {\bf 48} 4645

\bibitem{cahill} Cahill M E and  Taub A H 1971 {\it Comm. Math. Phys.}
                        {\bf 21} 1

\bibitem{lsode1} Hindmarsh A C 1983, in {\it Scientific Computing} 
                  Stepleman R S \etal (eds.) (North-Holland,
                  Amsterdam) 55

\bibitem{lsode2} Petzold L R 1983 {\it J. Sci. Stat. Comput.} {\bf 4} 136

\bibitem{rjlbook} LeVeque R J 1992 {\it Numerical Methods for Conservation
                          Laws} (Birkha\"user-Verlag, Basel)

\bibitem{rjl98}   LeVeque R J 1998 , in {\it Computational Methods for
         Astrophysical Fluid Flow,} 27th Saas-Fee Advanced Course Lecture
         Notes (Springer-Verlag, Berlin, to be published); also available at
         {\tt http://sirrah.astro.unibas.ch/saas-fee/}.

\bibitem{ibanez} Ib\'a\~nez J M$^{\rm {\underline a}}$,
                 Mart\'\i\  J M$^{\rm {\underline a}}$,
                 Miralles J A and
                 Romero J V 1992 {\it Approaches to Numerical Relativity,}
                (Cambridge University Press, Cambridge) p~223.

\bibitem{romero} Romero J V,
                Ib\'a\~nez J M$^{\rm {\underline a}}$,
                Mart\'\i\  J M$^{\rm {\underline a}}$
                and  Miralles J A 1996
                {\it ApJ} {\bf 462} 839

\bibitem{bgnc}  Brady P, Gundlach C, Neilsen D, and Choptuik M 1999 
                in preparation

\bibitem{gundlach_pc} Gundlach C 1999 personal communication

\bibitem{rob1} Roberts M D 1985 {\it Gen. Rel. Grav.} {\bf 17} 913

\bibitem{rob2} Roberts M D 1998 {\tt gr-qc/9811093}

\bibitem{gund1} Gundlach C 1998  \PR D {\bf 57} 7080

\bibitem{gund2} Martin-Garcia J M and Gundlach C 1998  {\tt gr-qc/9809059}

\endbib

\end{document}